# Adsorption and Diffusion of H atoms on $\beta$-PtO$_2$ Surface: The Role of Nuclear Quantum Effects

Yong Yang[1]* and Yoshiyuki Kawazoe[2,3]

1. *Key Laboratory of Materials Physics, Institute of Solid State Physics, Chinese Academy of Sciences, Hefei 230031, China.*
2. *New Industry Creation Hatchery Center (NICHe), Tohoku University, 6-6-4 Aoba, Aramaki, Aoba-ku, Sendai, Miyagi 980-8579, Japan.*
3. *Department of Physics and Nanotechnology, SRM University, Kattankulathurm, 603203, TN, India.*

**ABSTRACT:**

The adsorption and diffusion of H atoms on $\beta$-PtO$_2$(001) surface have been studied using first-principles calculations. The chemisorbed H atoms are found to bind preferentially on the top sites of O atoms due to the much larger adsorption energies with comparison to adsorption atop Pt atoms. The calculated energy barriers along the optimal diffusion paths are comparable with that of H diffusion on Pt(111). Within the WKB approximation, the nuclear quantum effects (NQEs) along the diffusion paths are investigated. It turns out that the NQEs are significant for the surface diffusion of H at room temperature and play a dominant role in cryogenic conditions.

--------------------------------------------------------------------------------

*Corresponding Author: Y. Yang (yyanglab@issp.ac.cn).

## I. INTRODUCTION

The adsorption and diffusion of hydrogen on solid surfaces play an important role in the applications of hydrogen-based energy resources: For instance, the hydrogen oxidation reaction (HOR) and/or hydrogen evolution reactions (HER) in proton-exchange membrane fuel cells (PEMFC), in which Pt-based materials are the commonly employed catalysts [1-3]. The processes involving the surface dynamics of hydrogen are also the central theme of heterogeneous catalytic reactions for the synthesis/decomposition of the hydrocarbons [4-7], hydrogen storage [8-16], as well as the hydrogen embrittlement phenomena in metals and alloys [16-20].

Compared to the tremendous researches on the properties of hydrogen adsorption on the surfaces of transition metals [1, 21], much less attention has been paid to the surfaces dynamics of hydrogen on transition metal oxides, which are of equal importance in technological applications including catalysis [22-25]. In this work, we study the adsorption and diffusion of hydrogen atoms on the (001) surface $\beta$-$PtO_2$, the dioxide of Pt which can be formed on the electrode surface of PEMFC during realistic applications [26]. In the anode reactions, the H atoms are striped of electrons, and transfer across the electrolyte (i.e., proton-exchange membrane) before arriving at the cathode in the form of protons ($H^+$). In the cathode reactions, the arrived protons are combined with the electrons which are transported through an electrical circuit, and react with the adsorbed $O_2$ molecules which are finally reduced to water. Although it is possible that the arrived electrons may combine firstly with $O_2$ molecules (and get the product $O_2^-$) and then react with $H^+$, the combination of $H^+$ with electrons would be the dominant when considering the much larger electron affinity of $H^+$ (~ 13.6 eV) with comparison to $O_2$ (~ 1.46 eV) [27]. That is, the initial step of cathode reaction takes place following the Volmer route [1, 28]: $H^+ + e^- \rightarrow H^*$, where the symbol * denotes the surface adsorption sites. The diffusion of the adsorbed H atoms on the electrode surface will therefore play a key role, by affecting the supply of H to $O_2$ and consequently the rate of oxygen reduction reaction (ORR), the rate-limited step in electrode reactions of PEMFC [1, 2]. This study is motivated by the fact of the existence of $\beta$-$PtO_2$ on electrode surface [26], and the necessity of understanding the diffusion process of H atoms on the surface of $\beta$-PtO$_2$, which is yet unclear to date.

Due to the light mass and low electron densities of H atoms, their quantum motions, or the nuclear quantum effects (NQEs) would be significant at room temperatures and below. It has been demonstrated theoretically and experimentally that the NQEs of H have significant influence on the structural and dynamical properties of condensed phases such as water [29-34], liquid hydrogen [35], and crystal polymorphs [36]. Accordingly, we will investigate the NQEs on the diffusion of H atoms on the $\beta$-PtO$_2$ surface.

## II. METHODS

The first-principles calculations were carried out using the VASP code [37, 38], which is based on density functional theory (DFT). The crystal of $\beta$-PtO$_2$ takes an orthorhombic CaCl$_2$-type structure [39], which is unique when compare to the other group-VIII metal dioxides that crystallize in the tetragonal rutile structure [40]. In our study, the $\beta$-PtO$_2$(001) surface is modeled by a six-layer slab, with a $p(2\times2)$ surface unit cell which repeats periodically along the *xy*-plane, separated by a vacuum layer of ~ 15 Å in the *z*-direction. The atomic positions of the bottom three layers are fixed to simulate the bulk state. More details about the structures of the $\beta$-PtO$_2$(001) surface can be found in a recent work [41]. The electronic structure calculations were performed using a plane wave basis set with an energy cutoff of 600 eV for the expansion of electron wave functions. The projector augmented wave (PAW) potentials [42, 43] were employed to describe the electron-ion interactions. The exchange-correlation interactions of electrons are described by the PBE type functional [44]. To make a comparison, the van der Waals interactions between H and the PtO$_2$ surface were described by using the van der Waals density functional (vdW-DF) [45, 46]. The vibrational frequencies of the adsorbed H atoms and the adsorption systems were computed using the density functional perturbation theory (DFPT) [47, 48]. For structural relaxation and total energy calculations of the H/PtO$_2$ system, a $4\times4\times1$ Monkhorst-Pack k-mesh [49] is generated for sampling the Brillouin zone (BZ), and a $2\times2\times1$ k-mesh is employed for the DFPT calculations. The total energy of a single H atom is calculated in a (12 Å × 12 Å × 12 Å) supercell with a 2×2×2 k-mesh and the spin-polarization effects included.

The adsorption energy ($E_{ads}$) of H atom is calculated via the following formula:

$$E_{ads} = E[PtO_2(001)] + E[H] - E[H/PtO_2(001)] + \Delta E_{ZPV} \quad (1),$$

where $E[H/PtO_2(001)]$, $E[PtO_2(001)]$, $E[H]$ are respectively the total energies of the adsorption system, the $PtO_2(001)$ substrate, and an isolated H atom. The term $\Delta E_{ZPV}$ is the energy correction due to the change in zero-point vibration energy of the H atom, from isolated state to surface adsorption state. In our calculation, $\Delta E_{ZPV}$ is computed as follows: $\Delta E_{ZPV} = \frac{1}{2}(\sum_{i,isolated} \hbar\omega_i - \sum_{j,ads} \hbar\omega_j)$, where $\omega_i$ and $\omega_j$ are respectively vibrational frequencies of H atoms at isolated and adsorbed state, and $\hbar$ is the reduced Planck constant. There is no vibrations for a single atom at isolated state, i.e., $\omega_i \sim 0$, and consequently one has $\Delta E_{ZPV} = -\frac{1}{2}\sum_{j,ads} \hbar\omega_j$.

To study the diffusion of H atoms on the $\beta$-$PtO_2$ surface, we employed the nudged elastic band (NEB) method [50, 51] implemented in VASP to locate the saddle points of potential energy surface, and search for the minimum energy path (MEP) of diffusion. With the determined energy barriers, we can study the diffusion dynamics of the adsorbed H atoms from one surface site to another. The nuclear quantum effects (NQEs) are investigated within the Wentzel–Kramers–Brillouin (WKB) approximation [52-54], in which the transmission coefficient for a particle tunneling through a potential barrier $V(x)$ is evaluated as follows:

$$T_r(E) = \text{Exp}[-\frac{2}{\hbar}\int_a^b \sqrt{2m(V(x)-E)}\,dx] \quad (2),$$

where $m$ is the particle mass, $\hbar$ is the reduced Planck's constant, and $E$ is the energy of the particle that is associated with motion in the $x$-direction. In our calculation, $E$ corresponds to the kinetic energies of the H atom due to thermal motions, and $m$ is the mass of H. The range of integration is: $a \leq x \leq b$, within which $V(x) - E \geq 0$.

## III. RESULTS AND DISCUSSION

**Adsorption of H atoms on $\beta$-$PtO_2$(001).** We begin with studying the adsorption of H atoms on $\beta$-$PtO_2$(001), surveying a number of plausible configurations (Figure 1), including the adsorption on the top sites of surface O and Pt atoms. The corresponding adsorption energies, zero-point energies and adsorption geometries are listed in Table I. We have compared the results obtained by spin-polarized and non-spin-polarized calculations, and found that the difference in adsorption energies is less than 0.1 eV, with the most stable configuration being the spin-unpolarized state. Therefore, the results obtained by non-spin-polarized calculations will be presented here.

For the configurations shown in Figure 1, the adsorption energies of the ones atop surface O vary from ~ 3.4 to ~ 3.5 eV, and that of adsorption atop Pt vary from ~ 2.5 to ~ 2.6 eV: All configurations are chemisorbed on the $\beta$-PtO$_2$(001) surface. When the van der Waals interactions are taken into account, the magnitude of adsorption energies differs by several tens of meV with comparison to the results given by PBE type functional. Meanwhile, the values of zero-point energy (ZPE) of H atoms, which are ~ 0.24 to 0.29 eV, keep nearly unchanged in the PBE and vdW-DF calculations. The H-O bond length of Configurations A to D is ~ 0.98 Å, and the H-Pt bond length of adsorption atop Pt is ~ 1.57 Å. Due to the stronger O-H bonds, the zero-point energies of adsorption atop O are larger by ~ 40 meV with comparison to the configurations atop Pt.

**Diffusion of H atoms on $\beta$-PtO$_2$(001).** Due to their much larger adsorption energies, the adsorption configurations atop O will have higher probability to be observed than the ones atop Pt. As a result, the diffusion processes of H atoms are more likely to take place among the top sites of surface O atoms. We have considered three typical diffusion paths: From Configuration A to B (labeled as Path I), Configuration A to C (Path II), and Configuration A to D (Path III), as illustrated in Figure 2. The energy profiles, together with some typical transition configurations along the three diffusion pathways are schematically displayed in Figures 3 to 5.

For the diffusion from Configuration A to B (Figure 3), the H atom needs to overcome an energy barrier of ~ 0.27 eV (labeled as $E_{b1}$) by rotating the orientation of the O-H bond formed with the substrate (transition Configuration IA), and then hop to the nearest neighboring surface O (transition Configuration IB). In the next step, the H atom arrives at a more stable Configuration IC, with the orientation of the O-H bond rotated again by ~ 90°, pointing towards a surface O nearby (right panel of Figure 3). The transition from Configuration IA to IC is spontaneous, due to the downhill characteristics of the potential energy surface. The transition of the diffused H atom from the intermediate Configuration IC to the final Configuration B involves hopping to the top site of the neighboring O (Configuration ID) and the rotation of the O-H bond, surmounting a barrier of ~ 0.21 eV (labeled as $E_{b2}$). As shown in Figure 3, the values of $E_{b1}$ and $E_{b2}$ are reduced to ~ 0.21 eV and 0.17 eV, respectively, when the van der Waals interactions are considered.

In spite of the higher energy barriers encountered, and the longer distances travelled, the diffusion from Configuration A to C (Figure 4) is similar to diffusion along Path I in several aspects: The rotation of OH orientations, the hopping of the H atom from the top sites of O to the nearest neighboring O, and the existence of a number of more stable configurations (compared with the starting and ending configurations) on the diffusion path. Along the diffusion Path I (Figure 3), the intermediate Configurations IB, IC, and ID are more stable than Configurations A and B. For Path II (Figure 4), the intermediate Configurations IIA, IID, IIE and IIF are also more stable than Configurations A and C. To make a comparison, the energy and geometric parameters describing the two most stable configurations found in Path I and II (Configuration IC and IIE) are also listed in Table I. Compared to the top sites adsorption configurations shown in Figs. 1(a)-(d), the angle (α in Table I) between O-H bond and the $β$-PtO$_2$(001) is much smaller, which is ~ 28.5 °(~ 14.1 °by vdW-DF) for Configuration IC and ~ 0° (~ 9.4 °by vdW-DF) for Configuration IIE. The largely tilted angle leads to effective hydrogen bonding interactions between O-H bond and the neighboring O atom (IC in Figure 3, IIE in Figure 4). The enhanced hydrogen bonding interactions are also reflected in the increased zero-point energies and the elongated O-H bond length of Configuration IIE (Table I).

From the results presented in Figures 3, 4 and Table I, one sees that van der Waals interactions have some nontrivial effects on the energy barriers encountered along the diffusion pathway, as well as the adsorption geometries of the transition states. As shown below, such modifications would play a role when the quantum tunneling effects of H are considered.

The situation is much more different for the diffusion along Path III (Figure 5), from Configuration A to D. Before arriving at the destination (Configuration D), the H atom starting from Configuration A has to overcome a much higher energy barrier (~ 2.2 eV), for which the van der Waals interactions can be neglected. The diffusion process involves the following steps: The reorientation of O-H bond (Configuration IIIA), the breaking of O-H bond and formation of Pt-H bond (Configuration IIIB), the breaking of Pt-H bond and diffusion across the underlying Pt atom to a neighboring O, resulting in a distinctly distorted surface Pt-O bond (Configuration IIIC), and finally the formation and reorientation of the new O-H bond (Configuration IIID and D).

The energy barriers shown in Figures 3 to 5 clearly indicate that the diffusion of H atoms along Path I or II is feasible while the diffusion along Path III is energetically prohibited. Such difference originates from the different bonding environment that the H atom experiences on the diffusion paths: In Path I or II, hopping from the top site of one O to another with the O-H bond kept and the underlying Pt-O bond nearly unaffected, while the diffusion in Path III involves hopping from the top sites of O to the Pt nearby with the breaking of O-H bond and the underlying Pt-O bond largely affected (the line shaped O-Pt-O angle is bended by ~ 17°). The large barrier in Path III is due to the different strength of O-H and Pt-H bonds (Table I), as well as the distortion of the underlying Pt-O bond due to H adsorption. More generally, it can be inferred that if *only* O atoms are in direct contact with the diffused H atoms, the energy barriers would be moderate or low, and the diffusion is energetically favored. Otherwise, if there are O and Pt atoms which interact alternatively with H atoms on the diffusion path, the energy barriers would be high and the process is unlikely to happen.

On the cathode of a PEMFC, the arriving H atoms react with the $O_2$ molecules with the initial step being H* + $O_2$ → OOH*. On bare Pt(111) surface which is commonly regarded as the prototype of an electrode surface, the diffusion barrier for an adsorbed H atom (H*) to react with $O_2$ is calculated to be ~ 0.25 eV [3]. Such a value is very close to the barriers ($E_b$ = 0.27, 0.21 eV by PBE; = 0.21, 0.17 eV by vdW-DF) for a single H atom diffusion on $\beta$-PtO$_2$(001) surface via Path I (Figure 3), and slightly lower than the barriers ($E_b$ = 0.39, 0.37 eV by PBE; = 0.39, 0.26 eV by vdW-DF) of diffusion via Path II (Figure 4). These results imply that the transport and supply of H atoms on the electrode surface *will not* be significantly affected even when the Pt(111) surface is switched to $\beta$-PtO$_2$(001) during the ORR reactions.

We note here that the calculations of MEP haven't included the solvation effects of the surrounding water molecules. It has been shown by previous works [55-57] that during the process of HER and HOR, the energetics and reaction dynamics of the H atoms on Pt electrode are almost unaffected by the interfacial water. This is originated from the small dipole moment of the adsorbed H in the surface normal direction and consequently the weak dipole-dipole interactions between water and H atoms. Similar situation may be expected in the H/PtO$_2$ system, due to the fact that the adsorbed H is nearly charge-neutral upon adsorption on the electrode surface, and the dipole moment is also small in the surface normal direction (~ 0.5 Debye for adsorption atop O). Furthermore, the diffusion mainly involves hopping among the top sites of surface O, for which the energy corrections due to dipole-dipole interactions of water and H largely cancel out each other and therefore would have minor effects on the energy barriers.

**The role of NQEs on H diffusion.** For the diffusion and/or reactions on surfaces at atomic scale, the rate constant is proportional to the probability of passing through the energy barriers encountered, which is usually expressed in the form of $e^{-E_b/(k_BT)}$, where $E_b$ is the height of energy barrier, $k_B$ is the Boltzmann constant and $T$ is temperature. In the realm of classical physics, a particle can overcome an energy barrier only when its kinetic energy is equal to or higher than the barrier. In quantum physics, the situation is much more different: A particle can pass through a barrier which is higher than its kinetic energy at non-zero probability. This is the so-called quantum tunneling phenomenon. At the thermal equilibrium state, the kinetic energy distribution function of a single particle at temperature $T$ is [58]:

$$p(E_k) = 2\pi(\frac{1}{\pi k_B T})^{\frac{3}{2}}\sqrt{E_k}e^{-E_k/(k_B T)} \qquad (3),$$

where $E_k$ is the kinetic energy. For a classical particle, the probability of surmounting an energy barrier with a height of $E_b$ is:

$$P_C = \int_{E_b}^{\infty} p(E_k)dE_k \qquad (4).$$

Substitution of $p(E_k)$ with the expression in Eq. (3) and the integral gives:

$$P_C = \left(1 - \text{Erf}\left[\sqrt{E_b/(k_B T)}\right]\right) + \frac{2}{\sqrt{\pi}}\sqrt{E_b/(k_B T)}e^{-\frac{E_b}{k_B T}} \qquad (5),$$

where $\text{Erf}[x]$ is the error function, and the other parameters have the usual meanings. The first term $(1 - \text{Erf}[\sqrt{E_b/(k_B T)}])$ vanishes quickly and can be neglected when $E_b \geq 2k_B T$. The second term resembles the exponential term of Arrhenius equation, the popular empirical formula for the evaluation of rate constant. When the quantum motions of a particle come into play, the probability of quantum tunneling is calculated as follows:

$$P_Q = \int_0^{E_b} p(E_k)T_r(E_k)dE_k \qquad (6),$$

where the two terms $T_r(E_k)$ and $p(E_k)$ are expressed in Eq. (2) and Eq. (3), respectively. For each given value of $E_k$, $T_r(E_k)$ can be evaluated numerically using the data of potential $V(x)$ ($\leq E_b$), which is obtained by DFT calculations. In the situation where the kinetic energy of the particle is lower than the energy barrier ($E_k \leq E_b$), the classical transport is prohibited ($P_C = 0$) and only quantum tunneling is possible. The total probability of passing through the barrier is therefore $P_{tot} = P_C + P_Q$. In the calculation of $P_C$ and $P_Q$, the quantum vibrational levels are neglected due to two reasons: i) The diffusion on the surface is mainly due to the translational and rotational motions of H atoms, and ii) The typical excitation energy ($\hbar\omega_j$, twice the zero-point energy listed in Table I) between the quantum vibrational levels is ~ 0.5 eV, which can hardly be activated at room temperature conditions, and the H atoms mostly stay at their zero-point vibrational state, which does not contribute to the diffusion dynamics. From Eq. (2), one sees that the term $T_r(E_k)$ is nontrivial only when the particle mass is small, which explains why the nuclear quantum effects (NQEs) are more pronounced in hydrogen-involved systems than other systems which consist of heavy atoms. Increase of particle mass leads to exponentially decrease of the tunneling probability.

For a H atom that diffuses on $\beta$-PtO$_2$(001) at room temperature ($T \sim 300$ K), the classical and quantum tunneling probability, and their sum of surmounting the energy barriers along the three paths are summarized in Table II. The results obtained by PBE and vdW-DF calculations are listed for Path I and II, to show the effects of van der Waals interactions. For the three segmental diffusion paths A → IB, IC → B, and IIE → C, the classical and quantum tunneling probability are of the same order of magnitude. For the diffusion of IIA → IID, the classical probability $P_C$ is more than two times larger than the quantum tunneling probability $P_Q$. The underlying physical origin is: For a given temperature, $P_C$ depends only on the barrier height $E_b$, while $P_Q$ depends on the shape of barrier, including the barrier height and width, as well as the particle mass. To demonstrate this point, we go further to compare the two segmental paths of Path II: IIA → IID and IIE → C, which have similar height of $E_b$ (0.39 eV vs 0.37 eV). For the diffusion path IIE → C, $P_Q$ is slightly larger than $P_C$, while the order is inverse ($P_Q < P_C$) for IIA → IID, due to the larger barrier width in path IIA → IID than in IIE → C (Table II, 3.76 Å vs 1.87 Å). Based on the probability of surmounting the barriers (Table II), the reaction constant can be estimated by multiplying an attempting frequency of the adsorbed atom, which is the order of $10^{12}$ s$^{-1}$ [59, 60]. Approximately, the rate constants for the diffusion paths A → IB, IC → B, IIA → IID, IIE → C, and A → D are respectively $2.44 \times 10^8$ s$^{-1}$, $2.28 \times 10^9$ s$^{-1}$, $1.78 \times 10^6$ s$^{-1}$, $5.48 \times 10^6$ s$^{-1}$, and $1.36 \times 10^{-23}$ s$^{-1}$. It is evident that the diffusion along Path III (A → D) is dynamically forbidden. As shown in Table II, inclusion of the van der Waals interactions can lower the energy barriers by several tens of meV to $\sim 0.1$ eV (IIE → C). Addition of zero-point energy corrections to the potential energy surface will also cause minor modifications on the energy barriers (in the magnitude of several tens of meV) and consequently slight variations in the classical and quantum tunneling probabilities. However, the major results presented here are kept unchanged qualitatively.

To intuitively illustrate the effects of temperature and the role of NQEs, we have calculated the values of $P_Q$ and $P_C$ for the diffusion of H atom across the barrier along the segmental path A → IB (Figure 3). The results are shown in Figure 6, for temperatures ranging from 30 to 300 K. The value of $P_C$ drops rapidly with decreasing temperature. In contrast, the value of $P_Q$ varies much more smoothly with temperature, partly owing to the temperature-independence nature of the transmission coefficient $T_r(E_k)$. Another reason is the different range of integral: In present study the integral value of $p(E_k)$ within the specified range ($0 \leq E_k \leq E_b$) is always much larger than that obtained in Eq. (4). The role of van der Waals interactions is again demonstrated by the splitting data lines obtained from PBE and vdW-DF calculations. As indicated in Figure 6 and Table II, the NQEs contribute ~ 50% to the total probability of crossing the barrier (Path A → IB, Figure 3) at room temperature (~ 300 K). The NQEs will be even more important in cryogenic region, where the value of $P_Q$ is several to tens order of magnitude larger than $P_C$, which marks the dominant role of NQEs. On the other hand, the value of $P_C$ increases drastically with elevated temperatures and begins to play a nontrivial role at room temperature ($P_c \approx P_Q$) and beyond.

In the studies above, the three-dimensional (3D) energy profiles of diffusion are reduced to the MEP, the most probable pathway within the framework of classical statistical mechanics. With the reaction coordinates being described by a single variable (e.g., the distance traveled by H in our case), the MEP is mathematically equivalent to quasi-one-dimensional (1D) barriers and therefore can be treated by the WKB method. As a result, the diffusion processes along the other energy pathways, e.g., the ones with larger barrier heights while smaller widths which may bear similar values of $P_Q$ at low temperatures, have been neglected. Meanwhile, the dynamics of the underlying $PtO_2$ surface are not taken into account when evaluating the energy pathway of diffusion, which may also dramatically modify the magnitude of energy barriers along the MEP, especially at intermediate and high temperature conditions [61-63]. The shortcomings discussed here might have nontrivial effects on the results presented in this work, and will be the topic of future research.

## IV. CONCLUSIONS

In summary, we present an extensive study on the adsorption and diffusion of H atoms on $\beta$-PtO$_2$(001) using DFT calculations. The H atoms are chemically adsorbed on the $\beta$-PtO$_2$(001) surface, with the top sites of O atoms being energetically more stable than adsorption atop Pt. Strong hydrogen bond with substrate presents in the most stable adsorption configuration. The optimal diffusion paths are the ones along which the H atoms are transferred solely across the top sites of O atoms, with the energy barriers being comparable to that of diffusion on Pt(111) surface. Based on the energetic and geometric parameters determined by DFT calculations, the WKB approximation has been employed to investigate the NQEs, which are found to be overwhelmingly dominant in cryogenic conditions. The NQEs continue to play an important role in the surface dynamics of H even at ambient conditions. The present results demonstrate the significance of quantum motions of H atoms in surface diffusion and related reactions. Looking ahead, comparison with the results obtained by quantum dynamical methods such as the path integral molecular dynamics (PIMD) [34], which is too much time-consuming and still beyond our affordable computational resources, would provide more insights into the role of NQEs on surface-based processes.

**ACKNOWLEDGEMENTS**

This work is financially supported by the National Natural Science Foundation of China (Grant No. 11474285). We gratefully acknowledge the crew of Center for Computational Materials Science of the Institute for Materials Research, Tohoku University for their continuous support of the SR16000 and Cray XC50-LC supercomputers. We also thank the staff of the Supercomputing Center (Hefei Branch) of Chinese Academy of Sciences for their support of supercomputing resources.

**Table I.** Calculated adsorption energies ($E_{ads}$) and zero-point energies (ZPE) of H atoms on $\beta$-PtO$_2$(001), and the H-O bond lengths ($d_{OH}$), H-Pt bond lengths ($d_{PtH}$), and the angles formed between the O-H or Pt-H bonds and the $\beta$-PtO$_2$(001) surface ($\alpha$). For each quantity, the results obtained by PBE and vdW-DF calculations are in the upper and lower lines, respectively.

| | A | B | C | D | E | F | G | H | IC | IIE |
|---|---|---|---|---|---|---|---|---|---|---|
| $E_{ads}$ (eV) | 3.436 | 3.529 | 3.427 | 3.537 | 2.576 | 2.493 | 2.576 | 2.489 | 3.649 | 3.788 |
| | 3.517 | 3.593 | 3.516 | 3.594 | 2.621 | 2.542 | 2.620 | 2.543 | 3.723 | 3.745 |
| ZPE (eV) | 0.285 | 0.284 | 0.287 | 0.287 | 0.242 | 0.239 | 0.242 | 0.239 | 0.289 | 0.293 |
| | 0.286 | 0.287 | 0.286 | 0.287 | 0.242 | 0.234 | 0.242 | 0.234 | 0.323 | 0.322 |
| $d_{OH}$ (Å) | 0.981 | 0.983 | 0.978 | 0.978 | --- | --- | --- | --- | 0.981 | 1.045 |
| | 0.978 | 0.978 | 0.979 | 0.978 | --- | --- | --- | --- | 0.991 | 0.998 |
| $d_{PtH}$ (Å) | --- | --- | --- | --- | 1.564 | 1.567 | 1.564 | 1.565 | --- | --- |
| | --- | --- | --- | --- | 1.571 | 1.572 | 1.569 | 1.572 | --- | --- |
| $\alpha$ (°) | 50.83 | 52.25 | 51.49 | 51.96 | 41.86 | 46.74 | 42.12 | 45.76 | 28.50 | 0 |
| | 50.02 | 51.71 | 50.48 | 51.42 | 43.02 | 46.14 | 42.83 | 45.42 | 14.09 | 9.36 |

**Table II.** The calculated classical probability ($P_C$), quantum tunneling probability ($P_Q$), and the total probability ($P_{tot}$) of surmounting the energy barriers along the paths of H diffusion on $\beta$-PtO$_2$ (001) at 300 K. $E_b$ is the height of barrier, and $w$ is the width of barrier (illustrated in Figure 3). For each path, the data obtained by PBE and vdW-DF calculations are in the upper and lower lines, respectively.

| Diffusion Path | $E_b$ (eV) | $w$ (Å) | $P_C$ | $P_Q$ | $P_{tot}$ |
|---|---|---|---|---|---|
| A → IB | 0.27 | 1.72 | $1.12 \times 10^{-4}$ | $1.32 \times 10^{-4}$ | $2.44 \times 10^{-4}$ |
|  | 0.21 | 1.67 | $1.02 \times 10^{-3}$ | $9.60 \times 10^{-4}$ | $1.98 \times 10^{-3}$ |
| IC → B | 0.21 | 1.95 | $1.02 \times 10^{-3}$ | $1.26 \times 10^{-3}$ | $2.28 \times 10^{-3}$ |
|  | 0.17 | 1.40 | $4.34 \times 10^{-3}$ | $3.15 \times 10^{-3}$ | $7.49 \times 10^{-3}$ |
| IIA → IID | 0.39 | 3.76 | $1.29 \times 10^{-6}$ | $4.93 \times 10^{-7}$ | $1.78 \times 10^{-6}$ |
|  | 0.39 | 3.68 | $1.29 \times 10^{-6}$ | $5.21 \times 10^{-7}$ | $1.81 \times 10^{-6}$ |
| IIE → C | 0.37 | 1.87 | $2.72 \times 10^{-6}$ | $2.76 \times 10^{-6}$ | $5.48 \times 10^{-6}$ |
|  | 0.26 | 1.63 | $1.62 \times 10^{-4}$ | $2.47 \times 10^{-4}$ | $4.09 \times 10^{-4}$ |
| A → D | 2.20 | 5.54 | $1.24 \times 10^{-36}$ | $1.24 \times 10^{-35}$ | $1.36 \times 10^{-35}$ |

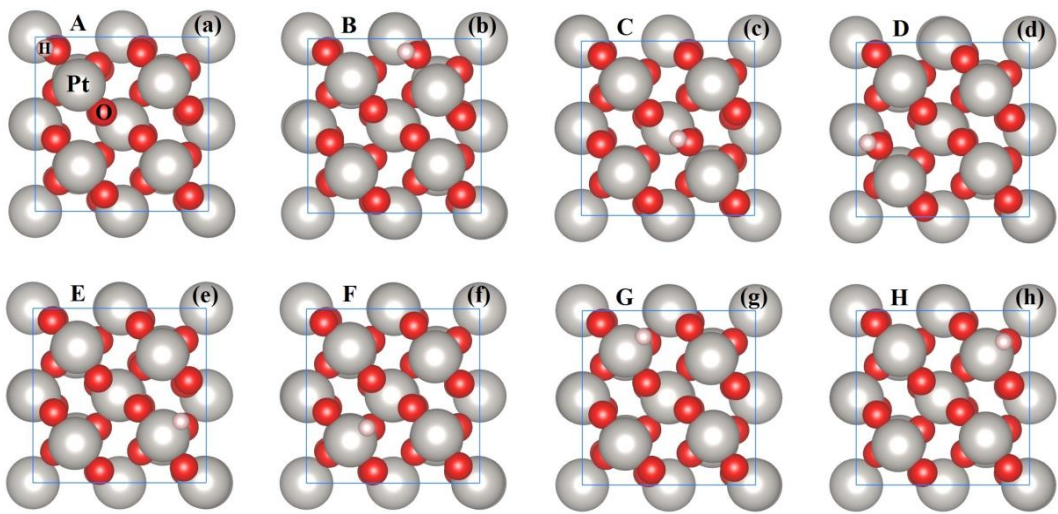

**Figure 1.** Top views of a single H atom adsorption on $\beta$-PtO$_2$(001), for the configurations atop O (panels **a – d**) and Pt (panels **e – h**), which are marked by capital letters A – H. The Pt, O, and H atoms are represented by silver (largest), red (second largest), and white (smallest) balls, respectively.

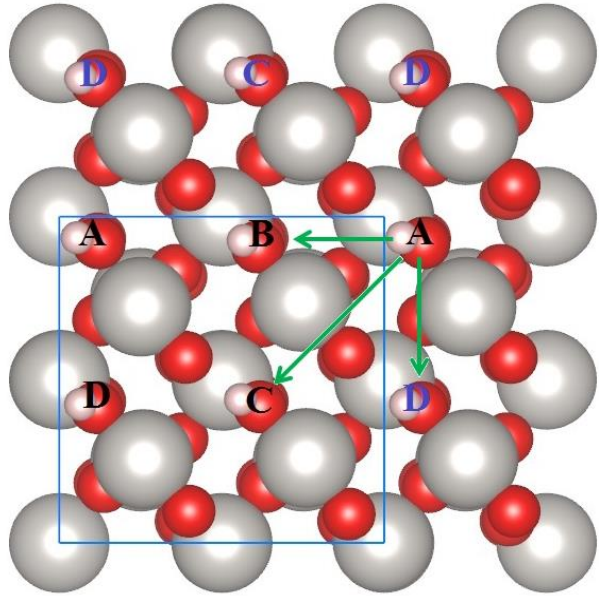

**Figure 2.** Schematics for the realistic diffusion of a single H atom, from Configuration A to Configurations B, C, and D, due to the constraint of periodic boundary condition (PBC). The supercell is marked, with the images due to PBC shown outside.

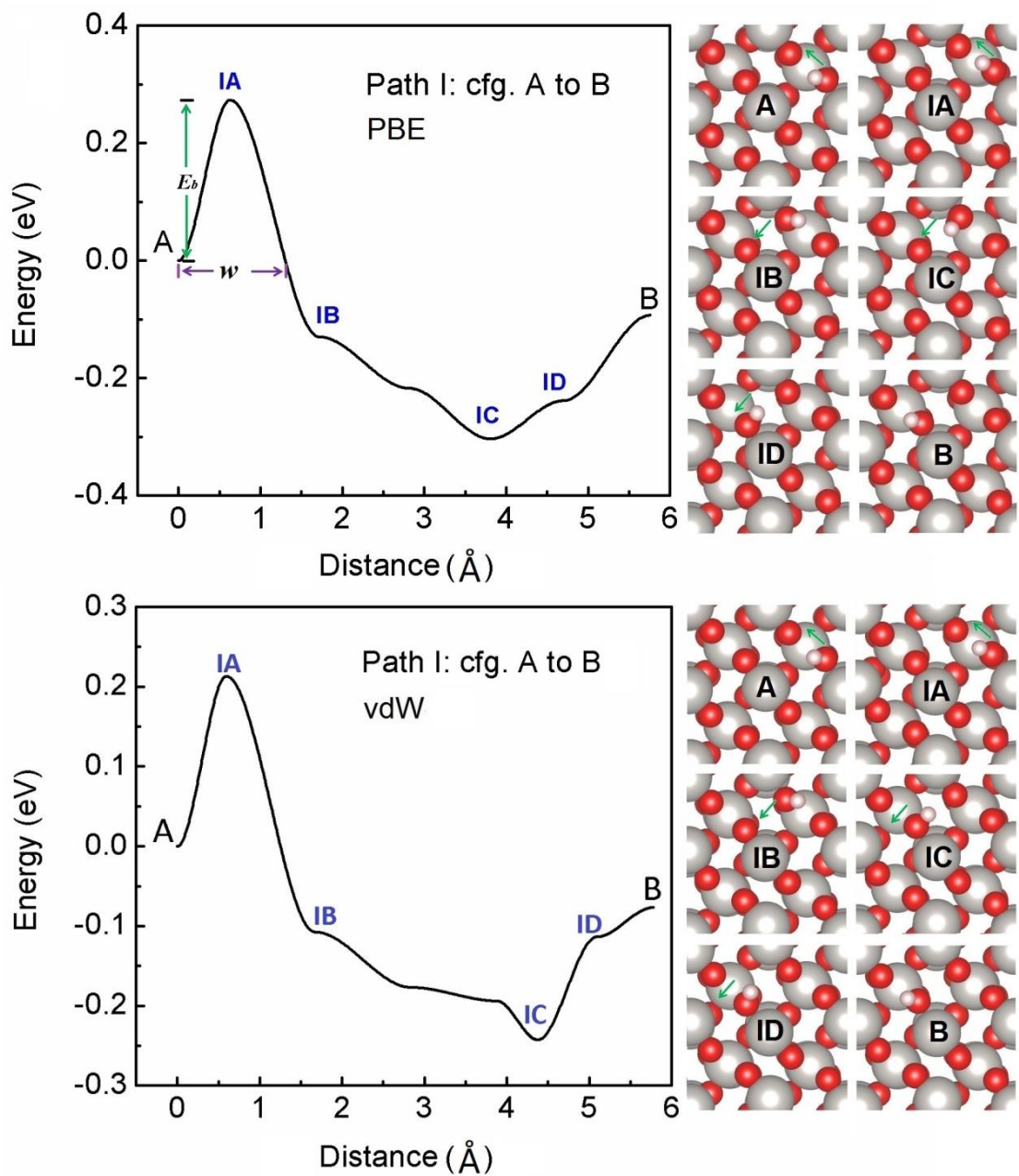

**Figure 3. Left panels:** The minimum energy path (MEP) for the diffusion of a single H atom from Configuration (abbr.: cfg.) A to B (Path I), with the typical transition configurations (IA, IB, IC, ID) marked. Horizontal axis presents the distance traveled by the H atom along the diffusion path. **Right panels:** Schematic diagrams for the starting, transitional and final configurations of H diffusion along Path I. The instantaneous moving directions of the H atom are indicated by arrows. The results calculated by PBE and vdW-DF type functional are in the upper and lower panels, respectively.

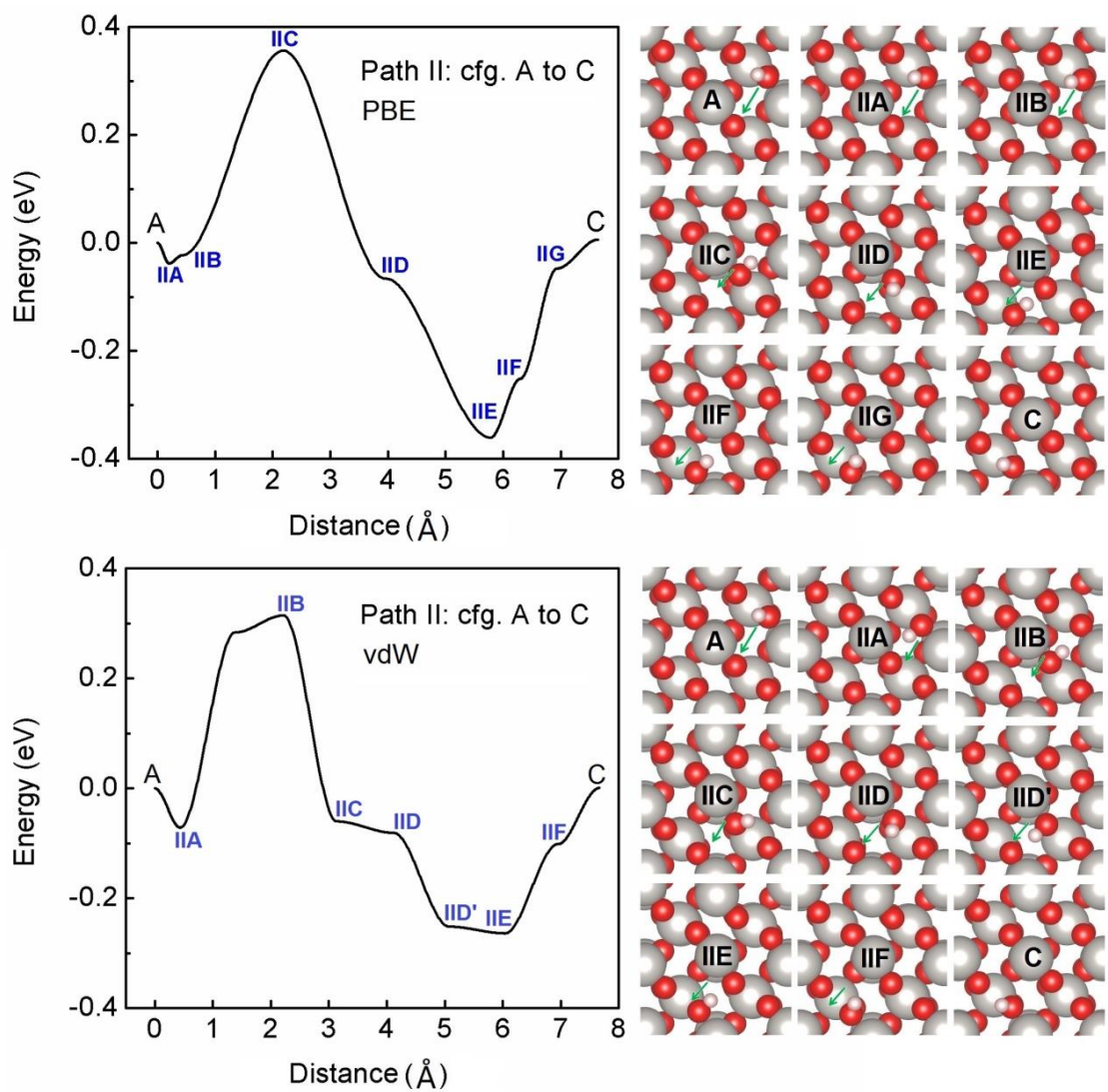

**Figure 4.** Similar to Figure 3 but for the diffusion of H atom along Path II (cfg. A to C).

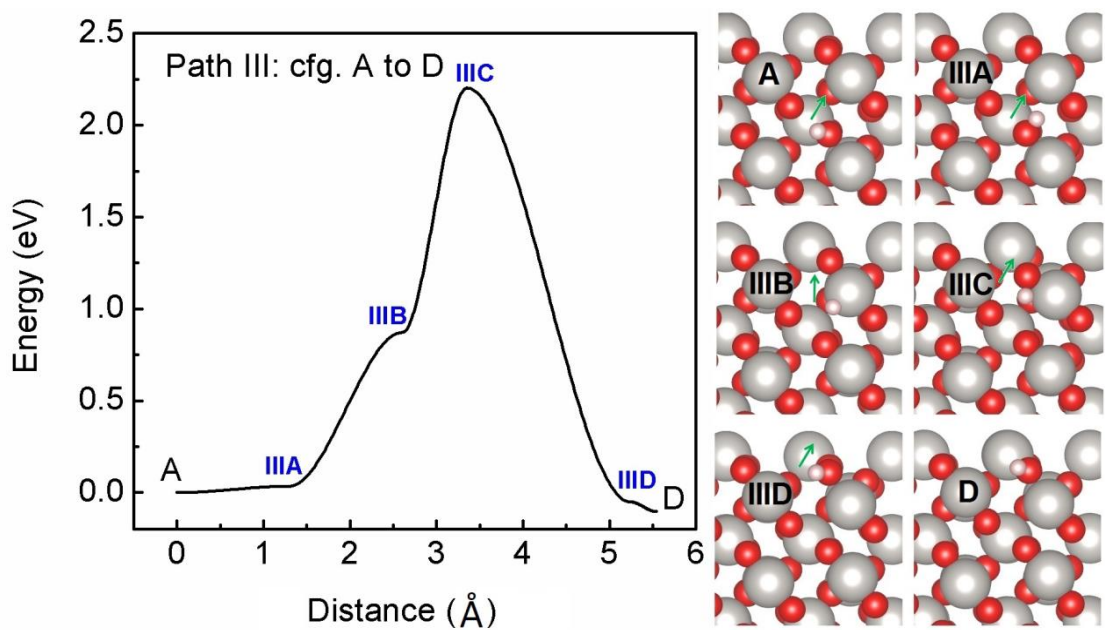

**Figure 5.** Similar to Figure 3 but for the diffusion of H atom along Path III (cfg. A to D), in which the calculations were done using PBE type functional only.

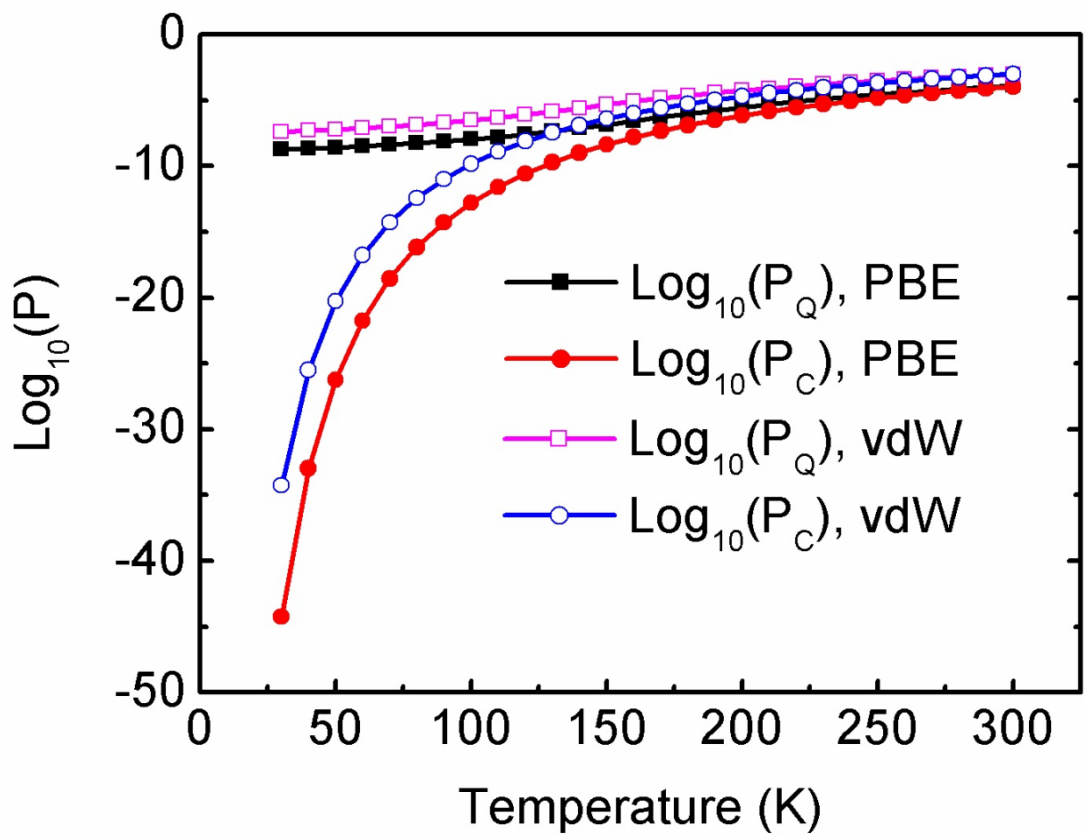

**Figure 6.** Logarithms of the classical ($P_C$) and quantum tunneling probability ($P_Q$), as a function of temperature, for a single H atom running across the energy barrier encountered along the segmental path cfg. A → IB, as illustrated in Figure 3. The results obtained by PBE and vdW-DF type calculations are shown for comparison.